\documentclass[a4paper,fleqn]{cas-dc}

\usepackage[authoryear,longnamesfirst]{natbib}

\begin{document}
\let\WriteBookmarks\relax
\def\floatpagepagefraction{1}
\def\textpagefraction{.001}

\shorttitle{Uniform probability in cosmology}

\shortauthors{S. Wenmackers}

\title [mode = title]{Uniform probability in cosmology}

\author[1]{Sylvia Wenmackers}[orcid=0000-0002-1041-3533]

\ead{sylvia.wenmackers@kuleuven.be}

\ead[url]{https://www.sylviawenmackers.be/}


\affiliation[1]{organization={Centre for Logic and Philosophy of Science, Institute of Philosophy, KU Leuven},
    country={Belgium}}

\begin{abstract}
Problems with uniform probabilities on an infinite support show up in contemporary cosmology. This paper focuses on the context of inflation theory, where it complicates the assignment of a probability measure over pocket universes. The measure problem in cosmology, whereby it seems impossible to pick out a uniquely well-motivated measure, is associated with a paradox that occurs in standard probability theory and crucially involves uniformity on an infinite sample space. This problem has been discussed by physicists, albeit without reference to earlier work on this topic. The aim of this article is both to introduce philosophers of probability to these recent discussions in cosmology and to familiarize physicists and philosophers working on cosmology with relevant foundational work by Kolmogorov, de Finetti, Jaynes, and other probabilists. As such, the main goal is not to solve the measure problem, but to clarify the exact origin of some of the current obstacles. The analysis of the assumptions going into the paradox indicates that there exist multiple ways of dealing consistently with uniform probabilities on infinite sample spaces. Taking a pluralist stance towards the mathematical methods used in cosmology shows there is some room for progress with assigning probabilities in cosmological theories.
\end{abstract}




\maketitle

\hbox{\vbox{\hyphenpenalty=50
    \begin{flushright}
    \begin{minipage}{77mm}
       \textit{``Infinite set paradoxing has become a morbid infection that is today spreading in a way that threatens the very life of probability theory, and requires immediate surgical removal.''---\citet[p.~xxii]{Jaynes:2003}}
    \end{minipage}
    \end{flushright}
}}
\vspace{1em}

The `infinite fair lottery paradox' refers to a formal incompatibility between the notion of a uniform probability measure on a countably infinite sample space (such as the set of natural numbers, which we take to be the positive integers) and the axioms of standard probability theory \citep[due to][p.~15]{Kolmogorov:1933}. One's reaction to this problem depends on the status one assigns to Kolmogorov's theory or the arguments that support it. If this theory---or an alternative that agrees on the two cases at hand---is taken to be a constraint for all our thinking about probabilistic scenarios---\textit{i.e.}, as the sole normative standard for reasoning about lotteries---then it seems to follow that the notion of a fair lottery on a countably infinite sample space is intrinsically paradoxical---that the very idea is mistaken or meaningless. This is the verdict reached by \citet{Jaynes:2003}, \citet{Guth:2018}, and \citet{Norton:2021}.

However, if one starts with a different theory as the normative standard for reasoning about lotteries or if one takes a pluralistic stance, allowing multiple formalisms for representing uncertainty (each with their own virtues and vices, and each with a certain range of applicability), then it may yet be possible to make sense of the notion of a fair lottery on a countably infinite sample space. In fact, the problems that arise from combining the notion of a fair lottery on the set of natural numbers with the axioms of Kolmogorov's theory has been a source of motivation for developing alternative probability theories, in particular, by \citet{deFinetti:1974} and \citet{Benci-etal:2013,Benci-etal:2018}.

Recently, the problem has resurfaced in the context in cosmology, albeit with little connection to their treatment in the earlier literature on the foundations of probability. The issue is related to the `measure problem' in cosmology, which refers to ambiguities in the assignment of probabilities to events in an infinite sample space. Although the measure problem is a more general issue, which remains even if the aforementioned paradox can be circumvented, a specific case of assigning uniform probability measures on infinite phase spaces has popped up in models of eternal inflation.

The problem occurs in the context of inflation theory, which posits that our universe underwent a rapid expansion relatively shortly after the Big Bang. Inflation theory has been very influential in cosmology in the past decades, leading many authors to comment on this issue. In physics, for instance, by \citet{Guth:2000} and \citet{LindeNoorbala:2010}, and in philosophy of science, for instance, by \citet{Smeenk:2014}, \citet{Curiel:2015}, and \citet{Norton:2021}. Some physicists take it to be a serious problem for the theory \citep{Ijjas-etal:2013}.

The main goal of this paper is to make the connection between the debate in cosmology and the older discussion on uniform measures in the foundations of probability explicit. Section~\ref {sec:paradoxKolmogorov} reviews the axioms and some further definitions and key theorems of standard probability theory. A paradox arises when this theory is combined with the notion of a fair lottery on the set of natural numbers. Section~\ref{sec:AlternativeFormalisms} discusses four alternative formalisms that do not lead to inconsistencies when they are combined with the notion of a fair lottery on a countably infinite set. Section~\ref{sec:IFLparadoxinflation} introduces the debate on eternal inflation and discusses which role the paradox plays there. Whereas we could take the notion of a fair lottery on the set of natural numbers for granted as the (theoretical) target system in the previous sections, here a crucial issue is whether or to which degree this is a relevant toy model to study probabilistic aspects of the (physical) target systems in cosmology. Section~\ref{sec:Conclusions2} formulates conclusions.

\section{A paradox in Kolmogorov's probability theory\label{sec:paradoxKolmogorov}}
\citet{Kolmogorov:1933} solved the first half of Hilbert's sixth problem: developing an axiomatic theory for the mathematics of probability. Moreover, building on earlier measure theoretic treatments of probability (by authors such as \'Emile Borel, Maurice Fr\'echet, Paul L\'evy, Constantin Carath\'eodory, Johann Radon and Otto Nikodym), Kolmogorov managed to integrate probability theory into measure theory---a crucial strength that enabled his theory to become the standard mathematical theory for probability. This section first reviews Kolmogorov's axiomatic system, as well as some further definitions and theorems of the theory. Then, a paradox is considered that arise when Kolmogorov's theory is applied to a fair lottery on the set of natural numbers.

\subsection{Kolmogorov's probability theory\label{sec:KolmogorovProbabTheory}}
\citet{Kolmogorov:1933}'s probability theory assumes standard set theory (Zermelo--Fraenkel set theory with the Axiom of Choice) as the background theory. The axioms of probability theory are equivalent to the following. Let $\Omega$ be a non-empty set called the sample space, of which the elements represent elementary possible outcomes (atomic events). (In examples from physics, the sample space is often the relevant phase space or some subset thereof.) Let $\mathfrak{A}$ be a sigma-algebra over $\Omega$,\footnote{This means that $\mathfrak{A}$ is a collection of subsets of $\Omega$ that includes $\Omega$ and is closed under complement (so it includes $\varnothing$) as well as countable unions.} of which the elements represent arbitrary events. Let $P$ be a function $P: \mathfrak{A} \rightarrow \mathbb{R}$ for which the following axioms hold, called a probability function.

\begin{description}
    \item[(K1)] $P$ is \textbf{non-negative}: $P(A) \geq 0$ for all events $A$ in $\mathfrak{A}$.
    \item[(K2)] $P$ is \textbf{normalized}: $P(\Omega) = 1$.
    \item[(K3)] $P$ is \textbf{finitely additive}: $P(A_1 \cup A_2) = P(A_1) + P(A_2)$, for all mutually disjoint events $A_1, A_2$ in $\mathfrak{A}$ (\textit{i.e.}, $A_1 \cap A_2 = \varnothing$).
    \item[(K4)] $P$ is \textbf{countably additive}: $P \left( \lim_{n \rightarrow \infty} \cup_{i \in \{1,\ldots ,n\} } A_i \right) = \lim_{n \rightarrow \infty } \sum_{i \in \{1, \ldots ,n\} } P(A_i)$, for all countable families of mutually disjoint events $\{A_i\}$, with all $A_i$ in $\mathfrak{A}$ (\textit{i.e.}, $A_i \cap A_j = \varnothing$ whenever $i \neq j$).
\end{description}

If the above axioms hold, the triple $\langle \Omega, \mathfrak{A}, P \rangle$ is called a Kolmogorovian probability space. Axiom K3 is redundant since it is implied by K4. Nevertheless, it is mentioned explicitly for two reasons. First, the first three axioms are sufficient when $\Omega$ is finite (where K4 holds trivially) and second, an alternative theory below that drops K4 entirely will be considered.

Starting from a standard probability space  $\langle \Omega, \mathfrak{A}, P \rangle$, \textbf{conditional probability} can be defined by the following ratio formula (for non-null conditioning events, \textit{i.e.}, events which don't have probability zero):

\begin{description}
    \item $P(A_1 \mid A_2) =_\textrm{def} P(A_1 \cap A_2) / P(A_2)$, for all events $A_1$ and $A_2$ in $\mathfrak{A}$ where $P(A_2) \neq 0$.
\end{description}

A consequence of countable additivity (K4) and the ratio formula for non-null conditioning events is the law of total probability:

If $\{A_1, A_2, \ldots \}$ is a countable partition of $\Omega$ whose members are in $\mathfrak{A}$, with $P(A_i) \neq 0$ for all $i$ in $\mathbb{N}$, then $P(A)= \lim_{n \rightarrow \infty} \sum_{i \in \{1,\ldots,n\}} P(A \mid A_i) P(A_i)$, for all $A$ in $\mathfrak{A}$.

In many cases where $\Omega $ is infinite, atomic events are null events. (Consider, \textit{e.g.}, a uniform distribution on a closed real interval.) To cover such cases, Kolmogorov provided a more general definition of conditional probability. Standard probabilities conditional upon atomic null events are not defined as such: they are only defined relative to a sigma-algebra (and even then, non-uniquely). To this end, consider a sigma-algebra $\mathcal{S} \subseteq \mathfrak{A}$. Then, for any $A \in \mathfrak{A}$, we can define $P_\mathcal{S}(A \mid \cdot ): \Omega \rightarrow \mathbb{R}$ as any function such that:

\begin{description}
    \item $P_\mathcal{S}(A \mid \cdot )$ is $\mathcal{S}$-measurable and
    \item $P_\mathcal{S} (A \mid \cdot )$ is such that $P(A \cap S) = \int_{S \in \mathcal{S}} P_\mathcal{S}(A \mid \omega) d P(\omega)$, for all $S$ in $\mathcal{S}$.
\end{description}

The previous equality is called the \textbf{integral formula}, which generalizes the law of total probability. Unlike the law of total probability, the integral formula is not a theorem in Kolmogorov's theory, but a defining property of conditional probabilities, which is also applicable for all atomic and/or null conditioning events.

Let us briefly comment on this from the perspective of \citet[p.~xxii]{Jaynes:2003}, who thought that probabilities on infinite sets can only be well-defined when the relevant sets ``arise as well-defined and well-behaved limits of finite sets''.\footnote{\citeauthor{Jaynes:2003} took his cue from Gauss's view on infinity: ``a figure of speech, the true meaning being a limit'', as quoted by \citet[p.~451]{Jaynes:2003}. According to \citet[p.~485]{Jaynes:2003}, ``as long as we look only at the limit, and not the limiting process, the source of the error is concealed from view''. Similarly, \citet[p.~487]{Jaynes:2003} wrote: ``at least in probability theory, an infinite set should be thought of only as the limit of a specific (\textit{i.e.}, unambiguously specified) sequence of finite sets''.} Probabilities in Kolmogorov's theory indeed depend on limiting processes, in two ways.

The first way is directly apparent from the countable additivity axiom (K4). It was equally apparent in the original statement of the final axiom by \citet[p.~14]{Kolmogorov:1933}:

\begin{description}
    \item[(K4')] $P$ is \textbf{continuous}: if a decreasing sequence of events $A_1 \supseteq A_2 \supseteq A_3 \supseteq \ldots$ with all $A_i$ in $\mathfrak{A}$ such that $\lim_{n \rightarrow \infty} \cap_{i \in \{1,\ldots,n\}} A_i = \varnothing$, then $\lim_{n \rightarrow \infty} P(A_n) = 0$.
\end{description}

K4 and K4' are equivalent in the presence of axioms K1--K3. Yet another form of axiom K4 that is equivalent in the same sense \citep[as proven in][]{Benci-etal:2013} invokes the limit via a supremum:

\begin{description}
    \item[(K4'')] $P$ is \textbf{continuous}: $P(A) = \sup_{n \in \mathbb{N}} P(A_n)$, where $A = \lim_{n \rightarrow \infty} \cup_{i \in \{1,\ldots,n\}} A_i$ with all $A_i$ in $\mathfrak{A}$ and $A_i \subseteq A_{i+1}$ for all natural numbers $i$.
\end{description}

In the standard theory, also the following theorem holds \citep[for a proof, see][]{Benci-etal:2013}:
Conditional Probability Principle (CPP):

Let $\Omega_i$ be a family of events such that $\Omega_i \subseteq \Omega_{i+1}$ and $\Omega = \lim_{n \rightarrow \infty} \cup_{i \in \{1, \ldots ,n\}} \Omega_i;$ then, eventually $P(\Omega_i) > 0$ and, for any event $A$ in $\mathfrak{A}$, we have that $P(A) = \lim_{n \rightarrow \infty} P(A \mid \Omega_i)$.

So, the limit of conditional probabilities of an event relative to a suitable family of sets equals the unconditional probability of the event. This shows how standard probabilities depend on limiting processes, as Jaynes required.\footnote{Curiously, \citeauthor{Jaynes:2003} thought it is not necessary to add countable additivity as an axiom. To develop his objective Bayesianism, Jaynes started from the derivation of probabilistic principles by fellow physicist \citet{Cox:1946}. \citeauthor{Cox:1946} developed his principles to be relations of `reasonable credibility' that follow from a Boolean algebra on propositions and common sense. \citet[pp.~651--653]{Jaynes:2003} showed that when these principles are applied to sets, they conform with Kolmogorov's preamble as well as the first three axioms, (K1), (K2), and (K3). On the absence of a principle equivalent to axiom K4, \citet[p.~464]{Jaynes:2003} commented: ``As $n \rightarrow \infty$ it seems rather innocuous to suppose that the sum rule [equivalent to (K2)] goes in the limit into a sum over a countable number of terms, forming a convergent series; whereupon our probabilities would be called countably additive. Indeed (although we do not see how it could happen in a real problem), if this should ever fail to yield a convergent series we would conclude that the infinite limit does not make sense, and we would refuse to pass to the limit at all.'' In addition, \citet[p.~653]{Jaynes:2003}: ``We do not know how Kolmogorov was able to see the need for his axiom (4) of continuity at zero; but our approach, in effect, derives it from a simple requirement of consistency.'' Moreover, although \citet[p.~15]{Kolmogorov:1933} clearly stated that his continuity axiom was independent of the initial axioms (at least on infinite domains), \citet[p.~653]{Jaynes:2003} doubted this: ``Kolmogorov's axioms (3) and (4) appear to be closely related; it is not obvious whether they are logically independent.''}

Second, in Kolmogorov's theory, ``conditioning upon a null event is underdetermined. Even if we fix all unconditional probabilities, and even if we specify the null event upon which we wish to condition, we do not yet fix unique conditional probabilities'' \citep[p.~745]{Rescorla:2015}. Indeed, the general definition of conditional probabilities (including those conditional on null-events) involves further choices: a sub-sigma-algebra and a specific conditional probability function, where the latter choice is constrained but non-uniquely determined by the integral formula. \footnote{See, \textit{e.g.}, \citet[§4]{Rescorla:2015} for more details and discussion in relation to the Borel--Kolmogorov paradox; \citet[§15.7]{Jaynes:2003} discussed this paradox, too.}

\subsection{Infinite fair lottery paradox\label{sec:IFLparadox}}
Assume that $\langle \Omega, \mathfrak{A}, P \rangle$ is a Kolmogorovian probability space, so $P$ is a non-negative, normalized and countably additive function. Further assume (ex absurdo) that $P$ describes a fair lottery on the set of natural numbers, which is taken to mean at least three things:
\begin{itemize}
    \item the sample space $\Omega$ is $\mathbb{N}$;
    \item the sigma-algebra $\mathfrak{A}$ at least contains the singleton outcomes $\{n\}$ for all natural numbers $n$;
    \item all singleton probabilities are equal: \textit{i.e.}, $P(\{n\}) = P(\{m\})$ for all natural numbers $n$ and $m$. 
\end{itemize}
The above assumptions are jointly inconsistent, which can be seen as follows:
\begin{itemize}
    \item If we assign an equal but non-zero probability to all singleton outcomes (necessarily positive by K1), then the countably infinite sum over all singleton probabilities diverges. By the axiom of countable additivity (K4), this should equal $P(\Omega)$, which is 1 by the normalization axiom (K2). So, this option leads to an inconsistency.
    \item If we assign probability zero to all singleton outcomes, then the infinite sum over all singleton probabilities is zero. By the axiom of countable additivity (K4), this should equal $P(\Omega)$, which is 1 by the normalization axiom (K2). So, this option also leads to an inconsistency.
    \item There are no other options to assign to the singleton outcomes. (In particular, the assumption of a real-valued $P$ function in the preamble rules out assigning an infinitesimal probability to the singletons.)
\end{itemize}

We conclude that the notion of a fair lottery on the set of natural numbers (or any countably infinite set, for that matter),\footnote{There is not necessarily a problem with an uncountable sample spaces: \textit{e.g.}, a uniform probability distribution on $[0,1]$ (or any finite interval) is unproblematic, though positing uniform probability on $\mathbb{R}$ (which can be thought of as $[0,1[ \times \mathbb{Z}$) is equally paradoxical as on $\mathbb{N}$.} where fairness is understood as singleton uniformity, is inconsistent with Kolmogorovian probability theory.

This paradox was famously discussed by \citet{deFinetti:1974}. In his work on personalist probability, he took this as an important argument against adopting countable additivity as a postulate; see also \citet{KadaneOHagan:1995}. (We return to this in section~\ref{sec:FAP}.) Hence, the paradox has also been called the ``de Finetti lottery'' \citep{Bartha:2004}, as well as ``God's lottery'' \citep{McCallArmstrong:1989}. The issue has also been discussed in terms of drawing a random integer \citep{KadaneOHagan:1995}, so we might also call it the ``random integer paradox''.

\subsection{Infinite-set paradoxing and natural density\label{sec:InfiniteSetParadoxing}}
\hbox{\vbox{\hyphenpenalty=50
    \begin{flushright}
    \begin{minipage}{77mm}
       \textit{``Not only in probability theory, but in all mathematics, it is the careless use of infinite sets, and of infinite and infinitesimal quantities, that generates most paradoxes.''---\citet[p.~451]{Jaynes:2003}}
    \end{minipage}
    \end{flushright}
}}
\vspace{1em}

The paradox belongs to a wider category of problems associated with probabilities on infinite sets, such as the Borel--Kolmogorov paradox. The latter is not a formal inconsistency within the standard theory, but a vivid illustration of the dependence of conditional probabilities on limiting processes. What both problems share is the need for well-defined infinite limits.

Section~\ref{sec:KolmogorovProbabTheory} already discussed Jaynes's warning that probabilities of infinite sets can only be well-defined when they are defined in terms of unequivocally defined limits on finite sets. In this context, he also mentioned the lottery on $\mathbb{N}$ \citep[p.~xxii]{Jaynes:2003}: ``For example, the question: `What is the probability that an integer is even?' can have any answer we please in $(0, 1)$, depending on what limiting process is to define the `set of all integers' (just as a conditionally convergent series can be made to converge to any number we please, depending on the order in which we arrange the terms).''\footnote{He elaborated on this in \citet[pp.~671--672]{Jaynes:2003}.} Indeed, the question about the probability of a `random integer being even' is related to Galileo's paradox of infinity. \footnote{In his \textit{Dialogues Concerning Two New Sciences}, Galileo discussed whether there are equally many or fewer perfect squares compared to all integers; from two inconsistent assignments, he concluded that comparing infinite quantities is meaningless. See \citet{Mancosu:2009} for a discussion. \citet[p.~672]{Jaynes:2003} also connected the probabilistic version of the problem to Galileo's paradox.} Yet, it has a generally accepted answer in number theory.

Number theorists do study certain properties of the natural numbers with probabilistic methods \citep[see, \textit{e.g.},][]{Tenenbaum:2015}. They use the natural labelling on the natural numbers (the identity function) and the probabilities they use are uniform. This amounts to studying the natural density of subsets of the natural numbers within the entire set.

The natural density or asymptotic density on the natural numbers, $d$, is defined for those subsets $A$ of $\mathbb{N}$ for which the limit exists:
\begin{equation*}
d(A) = \lim_{n \leftarrow \infty} \#(A \cap \{1,2, \ldots ,n \})/n,
\end{equation*}
where $\#$ is the counting measure (finite cardinality). For example, for the subset of even numbers, the natural density is $1/2$; for finite sets it is 0 and for co-finite sets it is 1.
This measure is real-valued, non-negative, and finitely additive. It is also normalized by construction. Yet, it fails to be a ready-made standard probability measure for a fair lottery on the natural numbers. There are two reasons for this failure: first, the subsets of $\mathbb{N}$ for which the measure is defined do not form a sigma-algebra;\footnote{The measure takes subsets of $\mathbb{N}$ as an input but it is not defined for all subsets. In particular, the limit can fail to exist for infinite sets that are not co-finite. Moreover, the sets on which this measure is defined do not form an algebra. The natural density measure can be extended to the full powerset of the set of natural numbers, albeit not uniquely, using the Banach limit. We can consider a set of extended probability measures, which gives us ranges of probability values for the sets that were unmeasurable on the first approach. The width of the interval can be regarded as a measure of how pathological a set is: it can be maximal (\textit{i.e.}, equal to 1), but it can also be much lower. So, mathematicians do gain something by investigating this option. At the same time, it seems unlikely that pathological sets will be of much interest to physicists, since it goes beyond what they can measure. Alternatively, a unique measure can be extended to a larger domain, for instance by what \citet{KerkvlietMeester:2016a} called ``weak thinnability'', but this still fails to be a sigma-algebra. } second, the measure is not countably additive \citep{KadaneOHagan:1995}.

The main take-away of this section is twofold. On the one hand, all applications of probability to infinite sample spaces require an explicit choice of a limit process (also called a method of regularization). The choice involves establishing an order for the elements and a sequence of cut-offs (or truncation). Since the natural numbers form an ordered set, $(\mathbb{N}, \leq)$, a natural choice is to order the elements by this native linear order and to place the cut-offs by the same order, which leads to an inclusive order on the family of initial segments. Indeed, this is the choice on which the natural density is based: it studies the limit of a sequence of finite fair lotteries on sample spaces of the form $\{0, \ldots, n\}$. It may be uniquely well-motivated by its naturalness and simplicity, but it is still a choice. Relative to another choice, the relevant limits will come out differently, especially for infinite sets that are not co-finite (such as the subsets of even numbers) as \citeauthor{Jaynes:2003} pointed out. \citet[§5]{Bartha:2004} illustrated this vividly in terms of his ``re-labelling paradox''.

On the other hand, the paradox that we have seen above is not fully resolved by specifying such a limiting process. So, something more is at stake in the specific case of a uniform probability measure on a countably infinite sample space. The paradox shows that asking for a real-valued, normalized, countably additive measure that is singleton-uniform over a countably infinite sample space is simply asking too much: there are no such measures; the conjunction of requirements is inconsistent.

This leaves open the possibility of a measure that is singleton-uniform over a countably infinite sample space by dropping one or more of the other constraints, although it is up for debate whether the resulting measure can still be interpreted as a probability. (Moreover, we will have to remain mindful of \citeauthor{Jaynes:2003}'s warning about the label-dependence of limits.) Four such proposals are reviewed in section~\ref{sec:AlternativeFormalisms}. Of course, it is also possible to stick to the standard formalism and drop the uniformity requirement, thereby changing the target system. This is the line taken in cosmology by \citeauthor{Guth:2018}, in his joint work presented in \citeyear{Guth:2018}. Yet, it would be incorrect to claim that an infinite lottery cannot be modelled probabilistically at all \citep[as also acknowledged by][]{GuthVanchurin:2011}. It merely cannot be done within the constraints of Kolmogorov's theory: that is the main conclusion from the paradox.

\section{Alternative formalisms\label{sec:AlternativeFormalisms}}
There are many proposals for representing uncertainty that can capture the notion of a fair lottery on a countably infinite set. I do not cover them exhaustively but focus on those four that are most relevant for our present purposes.\footnote{This is not a principled choice but merely sets some boundaries on the current exploration. Alternatively, we could opt for a theory that is purely qualitative (a comparative ordering, rather than a quantitative measure, which is not a total order and which is non-Archimedean in the case of a countably infinite and uniform lottery) or one that only assigns values to conditional probabilities, or at least takes those as basic (such as Popper functions). Or we can use lexicographical theory or one that is both qualitative and conditional. Or a decision theory, that also includes utilities.}

The first two alternatives have been motivated, at least in part, precisely by the inability of the Kolmogorovian theory to represent a uniform probability distribution on the natural numbers. The first such alternative is de Finetti's merely finitely additive probability (FAP) theory \citep{deFinetti:1974}; the second one is \citeauthor{Benci-etal:2013}'s non-Archimedean probability (NAP) theory \citep{Benci-etal:2013}. In the context of the measure problem associated with eternal inflation, \citet{Vanchurin:2015} concluded that the axiom of countable additivity is the root of the problem. He agreed that ``the possible modifications of the classical probabilities deserve to be explored further'', mentioning FAP theory explicitly and NAP theory in a footnote, although he did not follow up these suggestions. Two further approaches that are relevant for this paper are Norton's infinite lottery logic (ILL) \citep{Norton:2021} and a pragmatic approach (rather than a well-developed formalism) by physicists to drop the requirement for normalizability \citep{Goldstein-etal:2016}.

\subsection{FAP: de Finetti's merely finitely additive probability theory\label{sec:FAP}}
De Finetti (\citeyear{deFinetti:1974}) developed probability theory in a personalist context: as rational constraints on individual probability assignments (called coherence), while the numerical values are not uniquely determined and may differ between coherent individuals. De Finetti insisted that probability measures should be defined on all subsets of the sample space.\footnote{See \citet{Bingham:2010} for more on the history, including the role of L.~J. Savage.}

His merely finitely additive probability (FAP) theory is identical to Kolmogorov's except that it drops the countable additivity axiom (K4) and requires that the sigma-algebra $\mathfrak{A}$ equals $\mathcal{P}(\Omega)$. Without the latter, it is a strictly weaker (\textit{i.e.}, more permissible) theory, which allows Kolmogorovian probability spaces as a special case.

Assume that the triple $\langle \Omega, \mathfrak{A}, P \rangle$ obeys the three axioms of FAP, which I call a FAP space. It is now consistent to assume that $P$ describes a fair lottery on $\mathbb{N}$, with fairness again understood as uniformity over singleton outcomes. It suffices to take the natural density (from section~\ref{sec:InfiniteSetParadoxing}) and extend it to all of $\mathcal{P}(\mathbb{N})$ via a Banach-limit to find an example of such a FAP function. In this case, we find that $P(\{n\}) = 0$ for all natural numbers n. This implies, by the finite additivity axiom (K3) that $P(F) = 0$ for all finite subsets $F$ of $\mathbb{N}$ and $P(CF) = 1$ for all co-finite subsets $CF$ of $\mathbb{N}$. Intermediate probability assignments occur for infinite subsets of $\mathbb{N}$ that are not co-finite sets. For instance, the probability of the subset of even numbers is $1/2$. The probability of many other such sets is not uniquely determined (which was unproblematic in de Finetti's personalist context).

Perhaps a certain paradoxical `feel' remains. For instance, the connection between $P(\lim_{n \rightarrow \infty} \cup_{i \in\{1,\ldots,n\}} \{i\})=P(\mathbb{N})$ and $\lim_{n \rightarrow \infty}  \sum_{i \in\{1,\ldots,n\}} P(\{i\})$ has been severed (the former being 1 and the latter being 0). This is a direct consequence of the failure of countable additivity. The result may appear to be counterintuitive to those who have always worked within Kolmogorovian probability theory (or measure theory more generally). \citet[p.~465]{Jaynes:2003}: ``We are trying to make a probability density that is everywhere zero, but which integrates to unity. But there is no such thing [\ldots]''. Although there is no formal inconsistency in FAP theory, \citet[p.~466]{Jaynes:2003} thought the failure of countable additivity was paradoxical in its own right: ``The real issue here is: do we admit such things as uniform probability distributions on infinite sets into probability theory as legitimate mathematical objects? Do we believe that an infinite number of zeroes can add up to one? In the strange language in which these things are discussed, to advocate `finite additivity', as de Finetti and his followers do, is a devious way of answering `yes' without seeming to do so.''

Some authors find dropping countable additivity acceptable, since it was not well-motivated in the first place.\footnote{\citet[p.~15]{Kolmogorov:1933}: ``\textit{We limit ourselves, arbitrarily, to only those models which satisfy Axiom [(K4)].} This limitation has been found expedient in researches of the most diverse sort.''} Others have tried to give positive arguments for FAP theory. De Finetti (\citeyear{DeFinetti:1972,deFinetti:1974}) argued that if we have an intuition about infinite additivity at all, it would lead us to expect perfect additivity (\textit{i.e.}, including uncountable additivity), which we cannot have. On his view, demanding countable additivity is an arbitrary stopping point. Another important argument he gave was motivated by the fair infinite lottery case itself: he saw no principled reason why individuals shouldn't be allowed to assign probability uniformly over a countably infinite set of mutually exclusive possibilities \citep{DeFinetti:1972,deFinetti:1974}. In addition to de Finetti, FAP theory was embraced by other personalists \citep[in particular][]{Savage:1972}, as well as objective Bayesians \citep{Cox:1946}. In his review of de Finetti's arguments for FAP theory, \citet{Howson:2014} called the arguments ``compelling'' and his case against countable additivity ``unanswerable''. FAP theory has also been defended, for instance, by \citet{Seidenfeld:2001} in the context of Bayesian statistics; one of his motivations was the infinite fair lottery analysis by \citet{KadaneOHagan:1995}. \citet[pp.~323--324]{Kelly:1996} agreed that we may have to reject countable additivity, because if we stick to Kolmogorov's countably additive probabilities, we can only describe nonuniform distributions on the natural numbers (favouring initial segments), which has the consequence that we should expect to see a counterexample to a false universal hypothesis sooner rather than later. In the context of subjective probability, \citet[p.~17]{Easwaran:2014} proposed to use a FAP functions together with the order structure of the algebra of events to allow for more fine-grained decisions involving equiprobable events.

Many commentors, however, are less enthusiastic about dropping countable additivity, because standard measure theory assumes it (where it allows the use of series expansions, etc.)\footnote{Although finitely additive measure theory was developed, too: see, \textit{e.g.}, \citet{RaoRao:1983}.} and it is essential for the derivation of the strong law of large numbers.\footnote{It is not feasible to be exhaustive here, but some examples are \citet{Doob:1953}, \citet[§3.1]{Dudley:1989}, and \citet[p.~213]{Edwards:1995}.} \citet{Bartha:2004} agreed with dropping countable additivity for the infinite fair lottery, though he did not conclude in favour of FAP theory in other cases.

According to \citet[p.~xxi]{Jaynes:2003}, de Finetti's FAP theory opened up ``a Pandora's box of useless and unnecessary paradoxes,'' of which finite additivity itself is an example. For \citet[p.~466]{Jaynes:2003}, ```finite additivity' is a euphemism for `reversing the proper order of approaching limits, and thereby getting into trouble with non-normalizable probability distributions'.'' Moreover, he wrote \citep[p.~656]{Jaynes:2003}: ``Like Kolmogorov, de Finetti is occupied mostly with probabilities defined directly on arbitrary uncountable sets; but he views additivity differently, and is led to such anomalies as an unlimited sequence of layers, like an onion, of different orders of zero probabilities that add up to one, etc.''

If we take formal consistency as our standard, however, FAP theory suffices to illustrate that the notion of a fair lottery on a countably infinite sample space is not intrinsically paradoxical, but merely jointly inconsistent with the Kolmogorovian theory. Moreover, the fact that the standard theory is formally inconsistent with this notion may be an argument against its axiom of countable additivity, which was not sufficiently motivated in the first place, as de Finetti and others have argued. The other three alternatives allow us to model such lotteries, too.

\subsection{NAP: Benci et al.'s non-Archimedean probability (NAP) theory\label{sec:NAP2}}
De Finetti (\citeyear{deFinetti:1974}, Vol.~1: p.~118) claimed that countable additivity was not uniquely well-motivated: if anything, one would expect arbitrary additivity, including uncountable additivity. He took this to be unobtainable and settled for finite additivity as a result. However, it turns out to be possible to impose a different kind of infinite additivity, that is not limited to a particular cardinality (sometimes called `perfect additivity'). This approach has been developed, for instance, in the non-Archimedean probability (NAP) theory of \citet{Benci-etal:2013,Benci-etal:2018}, who were also motivated by the infinite fair lottery paradox.

Although his writing predates an axiomatic NAP theory, \citet{Skyrms:1983b} already pointed to the hyperreals as a possible way to imbue probability values with a memory. He was thinking of keeping a trace of past updates in the posterior, but for infinite lotteries hyperreal values fulfil a memory function already at the level of priors: if seen as the result of some limit process on real-valued functions, hyperreals tell us the standard value of that limit (the real part) as well as the rate of convergence (the infinitesimal part, which can represent linear convergence, quadratic convergence, logarithmic convergence etc.; this replaces de Finetti's ``different orders of zero''\footnote{\textit{Cf.} \citet{Lewis:1980}: ``I think these people are making a rounding error: they fail to distinguish zero chance from infinitesimal chance.''}). Let us explain why this may be exactly what is needed to remove the lingering sense of paradox in the FAP solution to the infinite fair lottery problem.

As \citet[p.~452]{Jaynes:2003} stressed: ``passage to a limit should always be the last operation, not the first.'' Likewise, \citet[p.~466]{Jaynes:2003} wrote: ``trying to pass to the limit at the beginning of a calculation can generate nonsense because crucial information is lost before we have a chance to use it.'' Yet, this is exactly what seems to happen in FAP theory: to construct a probability function on the powerset of $\mathbb{N}$, one first takes the infinite limit of $n$ over functions on the powerset of initial segments $\{1, \ldots, n\}$. If one later takes the countably infinite sum of singleton probabilities, which are all 0, one finds $\lim_{n \rightarrow \infty} \sum_{i\in\{1,\ldots,n\}} 0 = 0$. If one were somehow able to `delay' taking the limit on singleton probabilities, one would obtain $\lim_{n \rightarrow \infty} \sum_{i \in \{1, \ldots ,n \}} 1/n = \lim_{n \rightarrow \infty} 1 = 1$, which makes intuitive sense (since the singleton probabilities decrease in proportion to and simultaneously with the increase of tickets) and agrees with $P(\mathbb{N}) = 1$. The requirement of constructing a probability function blocks the option of delaying taking this limit first. Here, NAP theory offers a possible way out: it takes the (non-Archimedean) limit which `remembers' the convergence behaviour of the sequence of functions on the initial segments (with linearly decaying singleton probability). This ``crucial information'' is used when taking the (non-Archimedean) limit sum, such that the agreement with $P(\mathbb{N}) = 1$ is regained.\footnote{It seems that \citet[p.~347]{deFinetti:1974} would agree with this approach: ``It has been said that to assume that $0 + 0 + 0 + \ldots + 0 + \ldots = 1$ is absurd, whereas, if at all, this would be true if `actual infinitesimal' were substituted in place of zero.''}

To achieve this, NAP theory requires a departure from the Kolmogorovian theory that is more severe than for FAP theory. So far, we have assumed that probability functions take values on the set of real numbers, which are Archimedean. Archimedean means that any strictly positive probability value $p$ is such that there exists a natural number $n$ such that $p>1/n$. Unlike the standard real numbers, non-standard models of the real numbers (called hyperreal numbers, which are available in a non-standard model of real closed fields) are not Archimedean. This means that they include infinitesimals, \textit{i.e.}, numbers with an absolute value between 0 and $1/n$ for all natural numbers, $n$. So, the first difference occurs in the preamble, where NAP theory stipulates a different domain for its probability functions: the set of real numbers is extended to a non-Archimedean set. Like in FAP theory, the relevant sigma-algebra of a NAP function is always the powerset of the sample space.

NAP theory is defined in terms of four axioms, which have been designed to mimic the Kolmogorovian axioms. NAP theory is stricter in the first axiom, by requiring the (prior) probabilities assigned to non-empty events to be positive rather than merely non-negative; this requirement is known as strict coherence or regularity.

The main departure occurs in the fourth axiom, because countably infinite sums are not necessarily well-defined for non-Archimedean numbers. Instead of countable additivity, NAP theory stipulates a different infinite additivity axiom. This axiom includes a definition of a non-Archimedean limit, written as $\lim_{n \uparrow \alpha}$ for a real-valued series or function in $n$. This limit depends on a fine ideal (equivalent to a specific type of free ultrafilter, the existence of which depends on the Axiom of Choice) and is not uniquely defined (which is unsurprising in light of the non-uniqueness of real-valued FAP functions). In NAP theory, fixing such a limit requires choosing a directed set: a collection of the finite subsets on $\Omega$, such that the union of two sets in the collection is included in a third and the union of which covers all of $\Omega$. The directed set encodes which limiting processes determine the value of certain conditional probabilities \citep[see][§4 for details]{Benci-etal:2013}. In doing so, NAP theory heeds \citeauthor{Jaynes:2003}'s \citeyear{Jaynes:2003} warning on how to avoid paradoxes due to infinite sets. Although any finite sum of infinitesimal probabilities is necessarily infinitesimal, the non-Archimedean limit sum of infinitesimal probabilities may be zero, infinitesimal, or finite and non-infinitesimal (though at most 1)---depending on the specific values.

NAP theory is closely related to numerosity theory, a non-Archimedean theory for assigning sizes to sets \citep{BenciDiNasso:2003b}. For example, $num(\mathbb{N}) = \alpha$ is an infinite hypernatural number that represents the numerosity (\textit{i.e.}, `size', in a particular sense, different from cardinality or ordinality) of the set $\mathbb{N}$ (as well as linear divergence). This numerosity, $\alpha$, also appears in the notation of the non-Archimedean limit in alpha-theory \citep{BenciDiNasso:2019} and NAP theory. For example, the case of a fair lottery on $\mathbb{N}$ can be described by numerosities multiplied by the normalization constant, $1/num(\mathbb{N}) = \lim_{n \uparrow \alpha} 1/n = 1/\alpha$, which is a particular infinitesimal hyperreal number that encodes linear convergence to zero.

The description of a fair lottery on $\mathbb{N}$ within the context of NAP theory has been covered explicitly in \citet[§5.2]{Benci-etal:2013}. Rather than assigning zero probabilities to all singleton outcomes, as is the case in FAP theory, the NAP function assigns a particular infinitesimal to this possibility. Since the numerosity of a finite set equals its finite cardinality, we have that $num(\{n\}) = 1$ for every natural number n, and the singleton probability is $P(\{n\}) = 1/\alpha$. By finite additivity, the probability of finite sets equals their finite cardinality times $1/\alpha$. The non-Archimedean limit sum of all singleton probabilities equals unity (since $\lim_{n \uparrow \alpha} \sum_{i \in \{1, \ldots,n\}} P(\{i\}) = \lim_{n \uparrow \alpha} n/\alpha = \alpha/\alpha = 1$), which is the probability of $\mathbb{N}$. It seems, then, that no trace of the first paradox remains.

Let us briefly comment on the close connection between NAP theory and FAP theory. Like FAP functions, NAP functions are not uniquely determined: finite differences occur for the same sets, and infinitesimal differences are even more ubiquitous. For instance, according to NAP theory, the probability of the subset of even numbers is either $1/2$ or $1/2 - 1/(2\alpha)$.\footnote{The former is the non-Archimedean limit of the family of finite fair lotteries with an even number of tickets, the latter of those with an odd number of tickets. It is possible to fix one or the other by selecting an appropriate directed set.} If we round off the infinitesimal part of the assignments (on operation called `taking the standard part'), both assignments return ½, which is the same value as the FAP function (which is uniquely determined for this particular set). This holds in general and taking the standard part of a NAP function results in a FAP function. The stronger additivity property of the former is lost in the process.

The expected outcome of the fair lottery on $\mathbb{N}$ can be computed as the non-Archimedean limit of the arithmetic mean, \textit{i.e.}, $\lim_{n \uparrow \alpha} \sum_{i \in \{1,\ldots,n\}} P(i) i$, where each $P(i) = 1/\alpha$ and $\sum_{in \in \{1,\ldots,n\}} i = (n+1)n/2$. Hence, the expected outcome is $(\alpha+1)/2$, which is infinite. This is consistent with the observation that any finite number is too small as an estimate for the expected value, so the latter diverges.

\citet{Vanchurin:2015} suggested that it may be worthwhile to investigate NAP theory for applications in cosmology. At the same time, while NAP theory is connected to many other approaches,\footnote{See \citet{Wenmackers:2019a} for a review of the notion of infinitesimal probabilities, \citet{BrickhillHorsten:2018} for a representation theorem that connects NAP theory to Popper functions and to lexicographical probabilities, and \citet{ChenRubio:2020} for a related proposal in terms of surreal numbers (with utilities, which are not relevant here).} there is still debate in the literature on the foundations of probability about its acceptability, regarding non-uniqueness \citep[but see][for some replies]{Benci-etal:2018}, the theory's incompatibility with certain invariances \citep{Pruss:2021}.

\subsection{ILL: Norton's infinite lottery logic\label{sec:ILL}}
Another response to the infinite lottery paradox is \citeauthor{Norton:2021}'s (\citeyear{Norton:2021}) infinite lottery logic (ILL). Unlike the previous two, ILL is not an axiomatic formalism, but a reasoned approach for comparing events related to the specific case at hand. Moreover, ILL is not intended to be a quantitative theory of probability. Its purpose is instead to capture some constraints for cases where probabilities cease to apply.

In earlier work on the use of infinite systems in physics, \citet{Norton:2011} pointed out that the limit properties of a sequence of systems need not correspond with the properties of a limit system. Although he did not apply this analysis to the case at hand, it is relevant here: the limit of a sequence of fair lotteries on the initial segments of $\mathbb{N}$ (say, the natural density) need not agree with a fair lottery on $\mathbb{N}$---if the latter idealization exists at all.

Like cosmologist \citet{Guth:2000}, philosopher of science \citet{Norton:2021} concluded from the infinite fair lottery paradox that there is no probabilistic description of it tout court. Considering our conclusion at the end of section~\ref{sec:InfiniteSetParadoxing}, this response seems too strong, but it is a consequence of their stronger interpretation of `fair' as equivalent to full label-invariance (rather than merely singleton uniformity). As we will see in section~\ref{sec:measureprobleminflation}, Norton---like Guth---applied his ideas about uniform countably infinite lotteries to questions about inflationary cosmology, so this alternative approach is certainly relevant to include here.

Norton's ILL applies to some unspecified countably infinite set, which can be labelled, arbitrarily by the set of natural numbers. ILL posits a chance function $Ch$ that assigns values to all subsets of the initial set by referring to arbitrary subsets of the labels (the natural numbers) and takes values on the ordered set $$V = \{V_0, V_1, V_2, V_3, \ldots, V_\infty, \ldots, V_{-3}, V_{-2}, V_{-1}, V_{-0}\},$$ such that:
\begin{itemize}
  \item $Ch(F_n) = V_n$ for any finite subset, $F_n$, of the natural numbers with cardinality $n$;
  \item $Ch(I_n) = V_\infty$ for any infinite subset, $I_n$, of the natural numbers that is not co-finite;
  \item $Ch(CF_n) = V_{-n}$ for any co-finite subset, $CF_n$, of the natural numbers of which the complement has cardinality $n$.
\end{itemize}

The ordering on the value-set $V$ is taken to be some antisymmetric, transitive, and irreflexive order relation, $<$, such that: $V_0 < V_1 < V_2 < V_3 < \ldots < V_\infty < \ldots < V_{-3} < V_{-2} < V_{-1} < V_{-0}$. There is also an informal interpretation given to the values: $V_0$ is ``certain not to happen'', $V_n$ with $n>0$ is ``unlikely'', $V_infty$ is ``as likely as not'', $V_{-n}$ with $n>0$ is ``likely'', and $V_{-0}$ is ``certain to happen''.

Let us first show that ILL is not really fit for purpose to deal with the infinite fair lottery paradox (as defined in section~\ref{sec:IFLparadox}), before explaining what its merits are, which may make it relevant for applications in inflationary cosmology after all.

If we compare this proposal with FAP theory, on the one hand, we see that ILL can represent differences between finite sets as well as between co-finite sets, whereas FAP functions collapse both (on probability 0 and 1, respectively). On the other hand, ILL is unable to represent any differences among infinite sets that are not co-finite, even for those of which the natural density is uniquely determined (\textit{i.e.}, all FAP functions agree). This is a direct consequence of requiring full label-invariance and certainly intentional. Still, this approach seems to be conceding too much (but see section~\ref{sec:ILLagain}).

A comparison with NAP theory may be instructive, too. All NAP functions respect the partial order of set inclusion: opting for a total order requires throwing away some order information (as ILL does for all infinite sets that are not co-finite) or adding information (individual NAP functions). Neither seems adequate; see for instance \citet{Easwaran:2014} who criticized the addition of arbitrary probability assignments (to infinite sets that are not co-finite) in hyperreal credences---an objection that applies to NAP theory. There are alternatives: qualitative non-Archimedean theory \citep{DiBella:2018}, a FAP function together with the algebra of events for additional order information \citep[p.~17]{Easwaran:2014}, or an entire family of NAP functions \citep{Benci-etal:2018}.

Standard probabilities on infinite sample spaces are not fully label-invariant and require an additional choice of limit processes on finite sets to come out as well-defined (recall section~\ref{sec:InfiniteSetParadoxing}), even outside of the specific case of an infinite fair lottery. Requiring this strong sense of uniformity, as \citet{Norton:2021} did, precludes any treatment with probability measures or even the usual qualitative frameworks from the outset. So, ILL is no potential replacement or alternative for standard probability theory (nor was it intended to be).

Let us be clear on what Norton's ILL does achieve. \citeauthor{Norton:2021}'s (\citeyear{Norton:2021}) non-probabilistic ILL for countably infinite lotteries applies to a very underspecified problem: there is a countably infinite lottery that is uniform on singleton outcomes. It is not specified whether the sample space is $\mathbb{N}$, $\mathbb{N}\setminus \{1 \}$, $2\mathbb{N}$, $\mathbb{Z}$, $\mathbb{N} \times \mathbb{N}$, $\mathbb{Q}$ or any other countable set. Of course, we may label the elements of the target set by $\mathbb{N}$ in any case, but now the argument that the identity function on $\mathbb{N}$ and thus the family of initial segments (\textit{i.e.}, subsets of the form $\{1, \ldots, n\}$) are uniquely well-motivated as a starting point for defining the relevant limits (a strategy used in both FAP and NAP theory) fails, since the labelling by $\mathbb{N}$ was arbitrary to begin with.

Observe that the analogous question is already underspecified in the analogous finite case: if all that is given is that there is a fair finite lottery, you cannot define a unique probability function for it, because you haven't been told the number of tickets. This is not a reason to give up on using probability theory, since there may be more information to be gathered. It would be a natural approach to make an estimate of the size of the lottery and to build a higher-order model: a family of admissible probability functions, possibly with a probability measure over them. \citet{Norton:2021} is right to warn us that this higher-order model should not be accepted as the answer without additional empirical ground, but if this is taken as a toy example of a scientific problem and science is seen as dynamic, it makes sense to start building models even if they require more input than we currently have. After all, this may guide further empirical searches.

In any case, it seems misguided to regard ILL as a proper response to the infinite lottery paradox as specified in section~\ref{sec:IFLparadox}: with the (ordered) set of natural numbers as the sample space and with singleton uniformity as the constraint. Moreover, we should avoid conflating the general issue of label-dependence of probabilities on infinite sets (discussed in section~\ref{sec:InfiniteSetParadoxing}) with the specific issue of uniform measures on countably infinite sets. Yet, there is some evidence of confusion between those issues in the literature.

For instance, \citet[p.~568]{Guth:2000} claimed that a probability measure on an infinite sample space requires us to take the ratio of infinities, which he claimed to be ill-defined; likewise for \citet[§2]{Norton:2021}. In standard probability theory, the first claim is not true: it requires taking the limit of finite ratios (as made explicit by the Conditional Probability Principle in section~\ref{sec:KolmogorovProbabTheory}). In FAP theory, the natural density measure (extended to the powerset) is of this kind.\footnote{See also \citet[§20.4]{DorrArntzenius:2017}, who reply in the context of subjective probability (credence): ``this assumes that claims about proportions provide the only possible basis for favouring some credences over others''.} In NAP theory, it is straightforward to define the ratio of infinite numbers (since the hyperreals form a field). In both cases, there are rules on how (not) to define the limit.

Another piece of evidence for confusion is that both \citet{Guth:2000,Guth:2007} and \citet{Norton:2021} pointed out that different `regulators' (which both regulate the labelling order and the truncation of finite initial parts of sequences) can produce different limiting ratios: this is certainly correct (again, recall section~\ref{sec:InfiniteSetParadoxing}). They do not consider, however, that not all regulators are adequate for defining a probability measure.\footnote{Although \citet[§4]{Guth:2007} did not state explicitly that there is a canonical choice of the regularization method for the infinite fair lottery problem (based on the native ordering of $\mathbb{N}$), perhaps he admitted this implicitly, for he did go on to consider a preferred regularization method for the more vexing issue of an order of the pocket universes.} Section~\ref{sec:IFLparadoxinflation} returns to this, where we will be in a better position to understand Guth's claim and to appreciate Norton's proposal.

\subsection{Non-normalizable quasi-probability\label{sec:quasiprobab}}
Finally, I discuss a pragmatic approach to issues related with the internal inconsistency in a ``real-valued, normalized, countably additive measure that is singleton-uniform over a countably infinite sample space''. This approach is found, for instance, in \citet{Goldstein-etal:2016}. Like FAP theory, it only drops one of the assumptions of Kolmogorov's theory, in this case the requirement for normalization. By allowing non-negative yet non-normalizable (\textit{i.e.}, unbounded) measures and sticking to the other Kolmogorovian axioms, this proposal stays firmly inside the realm of standard measure theory. Admittedly, it is unusual to interpret a non-normalizable function as a probability: it is more common to simply call it a measure, but I employ the term ``quasi-probability'' here as well, to investigate how fruitful such an approach is. I will use the symbol $\mu$ rather than $P$ for such measures. For nearly all events $A$ it holds that $\mu(A) = \infty$, so defining its probability as $P(A) = \mu(A)/\mu(\Omega)$ does not work.

To make this proposal work, all assignments should be limited to a specific sample space and an associated sigma-algebra, just like in standard probability theory. Like FAP and NAP measures, non-negative, non-normalizable quasi-probability functions can always be extended to the full powerset of the sample space. The quasi-probability function takes non-negative values on the extended reals. Instead of normalization, any infinite sample space is assigned measure $+\infty$. One should not expect to be able to compare quasi-probabilities across different sample spaces (a situation which is not unlike that in NAP theory).

For a fair lottery on $\mathbb{N}$, the measure of a singleton is arbitrary, so we might take it to be 1 for simplicity. Then, the uniform measure on any finite set is equal to the finite cardinality of that set. The measure is countably additive, and indeed, the countable sum over all singleton quasi-probabilities diverges, as does the measure of $\mathbb{N}$. Therefore, it seems that the infinite fair lottery paradox leaves no aftertaste here.

In fact, on this measure, all infinite sets have measure $+\infty$. So, in terms of representing differences among infinite sets, this approach does worse than ILL. Whereas a FAP function discerns differences in measure between infinite sets that are not co-finite by assigning the same measure to all finite sets, a non-normalizable function discerns differences in measure between finite sets by assigning the same measure to all infinite sets. Only NAP functions can represent differences in both ranges at the same time. Section~\ref{sec:probabinflationaryST} will return to the severe limitations of this approach.

Giving up on normalization is not a common response in the probability literature,\footnote{A notable exception is \citet{Renyi:1955}, who developed an axiomatic theory that did not require normality. Instead, he relied on conditional probability functions to obtain results in the unit interval. The pragmatic approach by the physicists can be made rigorous by applying this theory.} but a pragmatic approach by some physicists. It is plausible that their choice is influenced by their curriculum: for many students of Physics, this proceeds from calculus and measure theory to probability theory. So, normalization is added last, and may therefore seem to be the easiest constraint to lift. We could call this the ``last in, first out'' heuristic.

In the hands of a different group, however, the same heuristic may lead to a different preferred solution. Those who start out by studying \citeauthor{Kolmogorov:1933}'s axiomatization (\citeyear{Kolmogorov:1933}), in which normalization is the second axiom and countable additivity the last one, may be more willing to let go of the latter. This may apply to the probabilists that we saw defending FAP theory in section~\ref{sec:FAP}.\footnote{Admittedly, pointing to the Physics curriculum cannot be a full explanation of the state of the debate, since Cox and Jaynes were physicists who defended FAP theory, and Guth is a physicist who sticks to Kolmogorov's theory by denying the admissibility of a uniform measure on countably infinite sets. Moreover, we will see in section~\ref{sec:measureprobleminflation}, that while many cosmologists take a non-normalizable measure for granted, they do demand more for probabilistic predictions.}

In any case, the ``last in, first out'' heuristic is not sufficient for a principled choice. The aim of this section was to open the discussion to show four available options, which show some of the alternatives to formalize the notion of a fair lottery on a countably infinite set. Let us now turn to the role our central paradox plays in contemporary cosmology and how the alternatives that avoid these inconsistencies may be applied there.

\section{The infinite fair lottery paradox in the context of eternal inflation\label{sec:IFLparadoxinflation}}
Although probability theory is a branch of pure mathematics, its real power lies in its connection to statistics, which has applications in nearly all branches of science. Most probabilists assumed that infinite sample spaces are merely a matter of idealization, since the relevant properties of the actual target system, the actual number of repetitions, etc. are all finite. For instance, \citet[p.~15]{Kolmogorov:1933} wrote: ``Infinite fields of probability occur only as idealized models of real random processes.'' This view leaves open the possibility to retreat to a finite model in cases where paradoxes arise, such as the ones discussed in the previous sections.

An important exception to the general assumption of `finiteness in reality' is the domain of cosmology: in this context, there are various physical dimensions and quantities that are candidates for potentially being unbounded, not merely in theoretical models but in reality; these include time, space, or the number of pocket universes. This blocks the evasive retreat to finite models and requires dealing with the paradoxes head-on, especially if there are reasons to suppose a uniform probability measure over the full range of at least one of those candidate parameters. As such, the problems from the foundations of probability theory may spill over to cosmology.\footnote{\citet[p.~673]{Jaynes:2003} warned that: ``In our view, this plague [of paradoxes in infinite-set theory] is far more serious than mere obscure language; it infects the substantive content of pure mathematics. [\ldots] For now, it is the responsibility of those who specialize in infinite-set theory to put their own house in order before trying to export their product to other fields. Until this is accomplished, those of us who work in probability theory or any other area of applied mathematics have a right to demand that this disease, for which we are not responsible, be quarantined and kept out of our field.'' \citet[p.~674]{Jaynes:2003} also quoted (Kline quoting) ``Gibbs on this subject: `The pure mathematician can do what he pleases, but the applied mathematician must be at least partially sane.' ''}

Various probabilistic paradoxes in cosmology have been discussed by cosmologists and philosophers. Here, I limit myself to the two paradoxes introduced in section~\ref{sec:paradoxKolmogorov}, which crucially involve a uniform measure on countably infinite sample spaces.

\subsection{The measure problem in cosmological models of inflation\label{sec:measureprobleminflation}}
Inflation theory \citep[originally proposed by][]{Guth:1981} is an addition to the standard Big Bang scenario, which postulates a brief period of accelerated expansion in the early universe, to account for---among other things---the uniformity of the observed microwave background radiation, while the current focus is on explaining specific small inhomogeneities in these data. \citep[See, \textit{e.g.},][for details.]{GorbunovRubakov:2011} It was soon discovered that inflation theories generically lead to so-called eternal inflation: which means that the exponential growth of regions with false vacuum is faster than the exponential decay from false vacuum to stability \citep{Guth:2000}. In other words: whereas locally inflation has stopped, it keeps going in other parts of the multiverse. Eternal inflation predicts an infinite multiverse of such regions, with possibly very different local properties. The regions in such a multiverse are called `pocket universes', `bubbles', or `island universes'. Due to the exponential nature of inflation, there is always room for more pocket universes. The second ingredient, superstring theory or M-theory is consistent with a huge number of false vacua and associated constants (including the vacuum energy that may be interpreted as the cosmological constant). \citet{BoussoPolchinski:2000} suggested that the theories for eternal inflation and M-theory are compatible, such that the possible false vacua from M-theory are realized in different pocket universes of the eternally inflating multiverse.

Meanwhile, inflation theory has become an important ingredient of standard Big Bang cosmology. However, since it became clear that eternal inflation leads to a multiverse, there is an ongoing debate on whether such a theory can ever be testable and if so, to what degree. Rather than engaging with this bigger issue directly, our aim here is more modest: to examine two instances of the lottery paradox that occurs in the context of inflation theory.

\citet[p.~126]{Smeenk:2014} characterized the measure problem in cosmology as a combination of two problems: first, the sample space has to be defined together with a physically motivated measure, $\mu$. Often, $\mu$ is not unique and not normalizable. Non-normalizability leads to the second problem: ``defining a probability distribution over the physical properties of pocket universes.'' Moreover, \citet[p.~126]{Smeenk:2014} pointed out that there are at least two instances of the measure problem in inflationary cosmology. The first instance pertains to ``[a]ttempts to estimate the `probability of inflation' '', where ``the ensemble consists of a set of solutions to EFE [Einstein's Field Equation] and the measure in question is the canonical phase space measure'' \citep[p.~126]{Smeenk:2014}. The second instance pertains to probability of physical properties (such as the value of the cosmological constant, $\Lambda$, and other properties) within an inflationary multiverse. In this case, ``the ensemble consists of a collection of observers (or some other type of object) occupying a single, connected multiverse'' \citep[p.~126]{Smeenk:2014}.

\subsection{Probability of inflationary spacetime\label{sec:probabinflationaryST}}
The first instance was discussed by \citet[p.~126]{Smeenk:2014} as well as \citet[§4.2]{Curiel:2015}, who related it to the probability of a natural number being even. The goal here is to determine the probability of inflation over solutions of Einstein's field equation.

The first step is to determine the sample space, in other words: to choose a suitable subset of solutions to Einstein's field equations. The usual choice consists of finite-dimensional truncations, an approximation to the infinite-dimensional phase space of Einstein's field theory, called the minisuperspace $\Gamma$: it contains Friedman--Lema\^itre--Robertson--Walker universes, which are homogeneous and isotropic solutions to the Einstein's field equations minimally coupled to a single homogeneous scalar field, $\Phi$. So, the event space is some algebra of subsets of this $\Gamma$.

The second step is to choose a suitable measure on $\Gamma$, to serve as a (proto-)probability measure on the event space. In statistical mechanics, it is customary to base typicality judgments on the uniform Lebesgue volume measure of the underlying Hamiltonian phase space. This choice is motivated by Liouville's theorem, which guarantees that this is a stationary measure, \textit{i.e.}, the Lebesgue measure is invariant under the dynamics as described by the Hamiltonian (though there may be other measures with this property). The Liouville measure on $\Gamma$ is the Lebesgue volume measure known as the Gibbons--Hawking--Stewart measure \citep{Gibbons-etal:1987}. While this measure is uniform by construction, it is not normalizable.

When the Gibbons--Hawking--Stewart measure of some event (represented by an element of the algebra on $\Gamma$) is finite, one may think of it as having zero probability in the sense of a FAP theory and as having an infinitesimal or zero probability in the sense of NAP theory. When the Gibbons--Hawking--Stewart measure of the event is infinite, but the measure of its complement is finite, one may think of it as having unit probability in the sense of a FAP theory and as having unit minus an infinitesimal or unit probability in the sense of NAP theory. So far, so good. However, when both the event and its complement have an infinite Gibbons--Hawking--Stewart measure, the result is indeterminate.

It might seem, then, that I arrive here at a proposal for accepting the Gibbons--Hawking--Stewart measure as a non-normalizable quasi-probability function, or perhaps a case to accept a proposal akin to Norton's ILL. However, both \citeauthor{Smeenk:2014} and \citeauthor{Curiel:2015} point out that we need more for probabilistic predictions, since many events have infinite measure, while their complement has infinite measure, too. (Recall that FAP and NAP functions do assign values in such cases.) To get a well-defined probability, based on the notion of an asymptotic density, this requires introducing a ``regularization procedure''.

Section~\ref{sec:InfiniteSetParadoxing} discussed the importance of fixing a specific limit process (in particular, choosing a nested family of finite subsets of the sample space). We have also seen that there exists a canonical choice for this in the case of the natural numbers: the identity function on the natural numbers leads to the family of initial segments, $\{1, \ldots ,n\}$ as the ``regulator''. Given the identity labelling and associated family of finite subsets, the limits of relative frequencies are well-defined and unique for a large collection of subsets of the sample space (though not on the full powerset). Number theorists use this in their definition of the natural density (section~\ref{sec:InfiniteSetParadoxing}) and it underlies both FAP measures (section~\ref{sec:FAP})\footnote{\citet{Sahlen:2017} has suggested that FAP theory (based on Cox's approach) may be preferable for applications in quantum cosmology over Kolmogorov's theory. In particular, \citet[p.~436]{Sahlen:2017} pointed to ``unaccounted-for correlations in the structure of global properties'' to motivate dropping countable additivity, since ``in principle the measure evaluated on the full sample space need not equal the sum of the measures of all disjoint subsets of the sample space. This can be understood to mean that integrated regions under such a measure do not represent probabilities of mutually exclusive states.'' However, it seems that this line of reasoning does not support FAP theory after all, since ``problematic negative probabilities and non-unitarity could also occur in this case.''} and NAP measures (section~\ref{sec:NAP2}). This canonical choice leads to a probability of $1/2$ for an even number.

\citet[p.~123 \& 126]{Smeenk:2014} introduced the analogy to the fair lottery on the positive integers to stress the importance of choosing extra structure on the sample space: a nested family of finite subsets that regulates the limiting behaviour of the resulting probability function. Rather than a solution to the measure problem, this \textit{is} the measure problem. In other words, the real issue of the measure problem in cosmology does not lie in the infinite fair lottery paradox, but in the correspondence between an infinite label set and the ensemble of physically relevant possible outcomes, which may not have any additional physically meaningful structure to allow for a canonical---or at least a uniquely well-motivated---choice.

\citet{GibbonsTurok:2008} and \citet{CarrollTam:2010} both accepted the motivation behind the Gibbons--Hawking--Stewart measure, yet still ended up with different probabilistic predictions, because they introduced different ways to remove the divergence. \citet{GibbonsTurok:2008} proposed to equate Friedman--Lema\^itre--Robertson--Walker universes that are so flat (\textit{i.e.}, with curvature below a certain cut-off) that they cannot be observationally distinguished from a perfectly flat universe (with exactly zero curvature). They showed that the resulting probability of inflation assigned to simple scalar field models is exponentially small and that this result is robust (\textit{i.e.}, it does not heavily depend on the chosen cut-off value).

\citet{CarrollTam:2010} agreed with the diagnosis of \citet{GibbonsTurok:2008} that the non-normalizability of the measure stems from flat universes. However, they disagreed about the appropriate response: \citet{CarrollTam:2010} regarded flat and nearly flat universes as physically meaningful solutions and did not accept empirical indistinguishability as a valid reason for removing almost all solutions. In fact, they took the fact that almost all Friedman--Lema\^itre--Robertson--Walker universes are flat (where `almost all' is measured by the Gibbons--Hawking--Stewart measure) as an important result, showing that the classical cosmological theory has no flatness problem.\footnote{Although the regularization method presented by \citet{CarrollTam:2010} was later found to be faulty and replaced by the work of \citet{RemmenCarroll:2013,RemmenCarroll:2014} who derived a measure (again by applying Liouville's theorem; this time on the space of flat universes), these conclusions still hold; an overview can be found in \citet[§4.3]{Carroll:2023}.} As \citet[§4.3]{Carroll:2023} explained, the flatness problem occurs by implicitly assuming a uniform measure on the curvature. However, what should be established is whether the physics is such that generic trajectories have initial conditions with very small curvatures or not. Since the Gibbons--Hawking--Stewart measure is uniquely well-motivated by the dynamics and it is proportional to a power of $-5/2$ of the curvature, it diverges at curvature zero. From this, Carroll and co-authors have concluded that nearly flat universes are generic rather than exceptional---a radically different conclusion from that of \citet{GibbonsTurok:2008}.

Since spacetimes tend to get flatter over time due to inflation, the discrepancy between the two approaches---the sleight of hand that introduces additional structure---can also be viewed as due to sampling universes at earlier or later times (see also: \citet[p.~126]{Smeenk:2014} and \citet[§4.2]{Curiel:2015}.

Admittedly, the conclusion of Carroll et al. can still be contested by contesting the Gibbons--Hawking--Stewart measure as authoritative. Although it has become customary in the literature to use Liouville-based measures to motivate typicality judgments, a worry for this approach is that the `choice' of the initial conditions need not depend on the dynamics at all. That they can come apart is clear from lab settings, where initial conditions are determined by a process external to the system under study. Of course, the situation in cosmology is different since it aims to model the physical world at the largest scale, such that there is nothing external to it. Yet, in general it is not the case that dynamical laws determine their own auxiliary conditions \citep[see, \textit{e.g.},][§4.3]{ThyssenWenmackers:2021}.

Recall, however, that also \citet{GibbonsTurok:2008} accepted the Gibbons--Hawking--Stewart measure as their starting point. Yet, building on the work of \citet{SchiffrinWald:2012}, \citet[§4.2]{Curiel:2015} came to the pessimistic conclusion that there is no physical justification for accepting it in cosmology. And even if the measure would be equally well-motivated in cosmology as it is in statistical mechanics, in the latter context it is commonly used only to arrive at typicality judgements, not probability assignments as such.

It should be clear, however, that the latter problem has little to do with the infinite fair lottery paradox: the situation here is more akin to not knowing either what the sample space is, or whether the distribution is uniform. Therefore, the partial conclusion for this first instance of the measure problem is that the infinite fair lottery paradox is of limited value since the cosmological problem runs deeper. Let us now turn to the other instance.

\subsection{Probabilities of pocket universes in an inflationary multiverse\label{sec:probabpocketuniverses}}
The second instance of a cosmological measure problem discussed by \citet[p.~126]{Smeenk:2014} is no less problematic. In this case, the goal is to start from a model of eternal inflation and to assign probabilities to properties (such as values of the cosmological constant, $\Lambda$) of the pocket universes. \citet{Guth:2007} wrote that in such models ``anything that can happen does happen, and it will happen an infinite number of times''. For some, this quote succinctly captures what \citet{Vanchurin-etal:2000} called the ``predictability crisis'' of inflation theory \citep[\textit{e.g.},][]{Ijjas-etal:2013}. For others, including Guth, it points to what is needed to overcome it: comparing the probability density of various types of events in pocket universes requires finding a well-motivated method for regularizing the infinities.

For three decades, physicists have been trying to find a suitable probability measure for models of eternal inflation. This has resulted in a long list of candidate measures, each with their virtues and vices.\footnote{See, \textit{e.g.}, \citet{LindeNoorbala:2010} for a partial overview. To give a rough indication of the debate: some measures are based on choosing a global cut-off of some variable, such as the proper time (which leads to a youngness bias) or the scale factor (which requires special tuning of a variable to avoid oldness and youngness problems), and considering the infinite limit of that cut-off. Other measures compare the pocket universes relative to a comparable time in their respective evolutions (called the stationary measure). Still other measures start from finite spacetime volumes at `intermediate times' inside pocket universes (known as the causal-diamond measure).} The fact that these measures yield different results may be interpreted in at least three ways. Firstly, it may be an illustration that searching for a measure is intrinsically futile because of the nature of the problem at hand. For instance, \citet{Ijjas-etal:2013} indeed feared that the failure to find a well-defined measure implies that inflation theory makes no empirically testable predictions whatsoever. They considered this ``multiverse-unpredictability problem'' to be so severe as to motivate the search for alternatives to inflation theories. Secondly, it may merely show that the current cosmological models are insufficient to identify the relevant order, which leaves hope for future developments. \textit{E.g.}, \citet{GuthVanchurin:2011} seemed to be optimistic. Thirdly, there may not be a real problem here at all: different conditional probabilities (associated with different regulators) may suggest different unconditional probabilities, which may merely appear to be incompatible, without leading to an outright inconsistency. After all, non-normalizable measures are only finitely conglomerable, which means that two infinite partitions may yield two disjoint ranges of probabilities conditional on all members of those partitions \citep[see, \textit{e.g.},][§4.19]{deFinetti:1974}. Countable conglomerability is a theorem of standard probability theory, but the proof crucially relies on the axiom of normality. Since we need to drop that axiom and use an alternative theory to represent a non-normalizable measure, disjoint conditional probabilities do not pose an additional formal problem.

Some aspects of this second measure problem in inflationary cosmology are also closely related to the infinite fair lottery paradox. Indeed, in his treatment of a model for eternal inflation, according to which infinitely many pocket universes are created, \citet[§6]{Guth:2000} \citep[as well as][§4]{Guth:2007} used an infinite fair lottery on the (positive) integers as a model system. Since there are infinitely many pocket universes, \citet[p.~568]{Guth:2000} wrote that ``[t]he fraction of universes with any particular property is therefore equal to infinity divided by infinity---a meaningless ratio. To obtain a well-defined answer, one needs to invoke some method of regularization.'' (Elsewhere in the paper, he also used the term ``truncation'' for this.)

Similar to the previous section, we see here again that Guth's emphasis was not on the fact that there exists no real-valued, normalizable, and countably additive measure that is singleton-uniform on a countably infinite sample space. Instead, he pointed to the more general issue that infinite limits are order-dependent, which leads to label dependence of probability measures (not just on countably infinite sample spaces). This is related to the observation that applying the principle of indifference across different parametrizations may lead to inconsistent results.

At this point, it seems that the introduction of an infinite fair lottery as a model system in inflationary cosmology is a red herring: the real issue at hand in both instances of the measure problem is not internal to the notion of an infinite lottery on the positive integers. Our main message here is that the random integer paradox does not automatically lead to an unsurmountable predictability problem for doing science in an infinite universe or multiverse. As a result, it is not futile for cosmologists to attempt to solve the measure problem in eternal inflation. The real pickle is in the external usage of the integers as labels (for spacetimes or pocket universes, respectively). In section~\ref{sec:ILL}, I flagged a potential confusion in \citet{Guth:2000} and \citet{Norton:2021}; we are now in a better position to appreciate their comments, not as claims about the infinite fair lottery paradox itself but about the lack of a canonical choice for the family of finite subsets on the sample spaces relevant in cosmology.

\citet[§6]{Guth:2000} acknowledged this much for the second instance and indicated that the actual problem is that pocket universes do not come with a preferred order: any observed order will be observer dependent. The core of the problem is identified by \citet[p.~569]{Guth:2000}:

   ``In the case of eternally inflating spacetimes, the natural choice of truncation might be to order the pocket universes in the sequence in which they form. However, we must remember that each pocket universe fills its own future light cone, so no pocket universe forms in the future light cone of another. Any two pocket universes are space-like separated from each other, so some observers will see one as forming first, while other observers will see the opposite. One can arbitrarily choose equal-time surfaces that foliate the spacetime, and then truncate at some value of t, but this recipe is not unique. In practice, different ways of choosing equal-time surfaces give different results.''

The fact that the foliation into spacelike hypersurfaces is observer dependent, since the pocket universes are in the absolute elsewhere of each other, need not be fatal to the project: one could consider the infinite family of all such measures and consider the interval of values that they allow collectively.\footnote{Approaches with families of (standard) probability functions have recently been explored under the term `imprecise probabilities'. \citet{Bradley:2019} gave an overview of fields that applied imprecise probabilities, including references to quantum physics \citep{SuppesZanotti:1991,HartmannSuppes:2010} and nonlinear models by structural model errors \citep{Frigg-etal:2014}. In addition, imprecise probabilities in the context of anthropic predictions in cosmology \citep[§3.3]{Benetreau-Dupin:2015}, the upshot being that the familiar Bayesian approach doesn't need to be left completely. The suggestion to consider imprecise non-Archimedean probabilities has first been made by \citet{Benci-etal:2018} but has not been applied to physics or cosmology yet.\label{fn:impreciseprobab}} By analogy to the Copernican principle, one could even assume that the limiting densities of physically relevant properties agree for almost all observers. This is the probabilistic equivalent of a typicality approach, and the latter is commonly used in statistical mechanics and cosmology. However, this does not solve the question of whether the measure should be defined in terms of the cut-off of t in the first place.

Moreover, even if we agreed that the fair lottery on the natural numbers was intrinsically problematic, we might take a step back and wonder why we even considered that as a relevant model system for eternal inflation in the first place. Cosmologists commonly assume that the set of pocket universes of the inflationary multiverse is countably infinite, but relatively little motivation is given for this stance. In an early paper, \citet{AryalVilenkin:1987} concluded that ``inflating regions [\ldots] form a self-similar fractal of dimension slightly less than 3''. For a more complicated inflationary model of a `recycling universe', \citet[§V.C]{GarrigaVilenkin:1998} reported a limiting fractal dimension of 3 for inflating regions and a fractal dimension of less than 3 for regions of true vacuum, corresponding to pocket universes. As \citeauthor[p.~2239]{GarrigaVilenkin:1998} also noted, ``The fractal structure of realistic models is of course more complicated.'' In any case, a non-zero but finite Hausdorff dimension is compatible with countable as well as uncountable infinity. (In fact, this is independent of standard set theory: it is uncountable if the continuum hypothesis holds.) So, it seems that we should take equally seriously the possibility that the collection of pocket universes in a fractal multiverse is uncountably infinite. On the other hand, \citet{Harlow-etal:2012} developed an infinite binary (or, in general, p-adic) tree model, which merely contains a countable infinity of nodes. Building on this model, however, \citet{Vanchurin:2015} concluded that the trajectories form an uncountable set that can be mapped to the real-valued $[0,1]$-interval and thus measured by the Lebesgue measure, thus escaping the formal inconsistency between countable additivity, normality, uniformity, and a countably infinite sample space.

\subsection{Another look at Norton's ILL\label{sec:ILLagain}}
While \citeauthor{Norton:2021}'s ILL (\citeyear{Norton:2021}) perhaps seemed strange in response to the fair lottery paradox, as discussed in section~\ref{sec:ILL}, it may fare better when viewed as response to the measure problem in inflationary cosmology. As we have seen, \citet[§6]{Norton:2021} argued that label invariance is mandatory.\footnote{He does this by defining the principle of mediocrity (by which we should expect to find ourselves in an arbitrary pocket universe) in a very strong way: not merely as equiprobability of singleton outcomes, but also requiring invariance under all relabellings.} A model for such a property indeed precludes any probabilistic measures when the outcome space is infinite, but it requires a separate argument to establish whether no natural labelling (or class of labellings) is available in cosmology. For instance, the time at which pocket universes are spawned may offer such a label, but this is just one suggestion among many and there may be a serious problem here, as we just saw.

\citet[p.~42]{Steinhardt:2011} gave an analogy for inflationary pocket universes with or without a property of interest: a hypothetical sack of coins containing two countably infinite collections of pennies and quarters. Whereas random sampling is informative for finite collections, \citeauthor{Steinhardt:2011} suggested this ceases to be the case for infinite collections. Since the cardinality of both collections is equal, we could sort them in many ways: \textit{e.g.}, one-to-one or such that each pile of ten pennies is matched to one quarter. \citet[§3]{Norton:2021} agreed with \citet{Steinhardt:2011} that neither of the methods of labelling coins is right. I agree, too, but this does not imply that no correct method exists.

Observe that all methods that are easy to describe require `peeking': one must observe whether the coin is a quarter or a penny before it can be labelled. For the resulting limiting relative frequency to be admissible as a probability measure, however, the regulator should not be of this type: which type of coin receives label $n$ should only depend on the types of coins that received labels 1 to $n - 1$, not $n$ itself.\footnote{This criterion is closely related to the work of von Mises on collectives (random sequences). The core idea was that for a sequence to be random, sub-sequences that are selected without `peeking' should result in a random sequence, too. \citet{Church:1940} gave the criterion for admissible place selection rules: it can be any recursive rule that uses the $n - 1$ first elements of the sequence to decide, whether or not the $n^\textrm{th}$ should be selected. So, admissible selections may only depend on the previous outcomes and the number of trials so far (not current or future outcomes). Although the question answered by von Mises and \citeauthor{Church:1940}presupposes a labelling, it can be applied to rule out inadmissible (re-)labellings, too. Reshuffling a sequence of pocket universes taking into account their properties (\textit{i.e.}, `peeking') does not in general leave the probabilities of their properties invariant.} In addition, the situation may come with a native labelling, such as temporal order. In the toy examples of the pennies and quarters, producing a sequence by blindly drawing without replacement would suffice. The situation in inflationary cosmology is certainly more complicated, due to the lack of experimental access and in addition there are different ways of taking limits. I do not want to dismiss these problems; rather, my aim is to disentangle the different issues and to point out that at least one of them---the issue of a uniform probability on a countably infinite sample space---is not as insurmountable as \citet{Guth:2000}, \citet{Steinhardt:2011}, and \citet{Norton:2021} take it to be.

In response to these issues, \citet[§6]{Norton:2021} asked rhetorically: ``Why demand that uncertainties be represented probabilistically when the background conditions speak against it?'' Thereby, he suggested that cosmologists should accept their failures of finding a unique probability measure (so far) as evidence that their model is one of the non-probabilistic kind, which admits fewer or weaker predictions than a probabilistic model. \citet{Norton:2021} showed how his ILL is relevant for cosmological models that have countably infinitely many pocket universes and suggested that cosmologists should accept that this is the strongest form of predictions their models have to offer.

Whether the expectation of further empirical ground is realistic in the case of an inflationary multiverse theory is a matter outside of the foundations of probability theory itself, which I leave to cosmologists to decide. Observe, however, that the background information could be such that even Norton's inductive logic is too strong. One example would occur when all we knew is that the creation of each pocket universe can be modelled as an infinite lottery on a countably infinite set: if we cannot assume that they are produced by one and the same such lottery, then even the ordering of the finite events is jeopardized. Another example would occur when we did not know which properties of the pocket universes are variable across different realizations. To be clear, these are merely possibilities; to the best of our knowledge, there are no such models in current cosmology.

Another problem that Norton has repeatedly flagged in earlier work and that appeared again in his (\citeyear[§2]{Norton:2021}) is that being equally uncertain or indifferent is not always captured best by a uniform probability function. A different context in which this worry arises is in forensics, where it has been argued that it may be better to opt for Dempster--Shafer conditioning and belief functions, rather than probability functions \citep{KerkvlietMeester:2016b}. At the same time, such approaches are formally connected to families of probability functions (\textit{cf.}\ footnote~\ref{fn:impreciseprobab} on imprecise probabilities), so it may be interesting to examine whether this approach is applicable to the measure problem in cosmology as well.

\section{Conclusions\label{sec:Conclusions2}}
\hbox{\vbox{\hyphenpenalty=50
    \begin{flushright}
    \begin{minipage}{77mm}
       \textit{``A paradox is simply an error out of control; \textit{i.e.} one that has trapped so many unwary minds that it has gone public, become institutionalized in our literature, and taught as truth.''---\citet[p.~451]{Jaynes:2003}}
    \end{minipage}
    \end{flushright}
}}
\vspace{1em}

Section~\ref{sec:paradoxKolmogorov} reviewed that a singleton-uniform, real-valued, and countably infinite probability distribution on an infinite support cannot be described by standard probability theory. There have been different responses to the associated paradoxes. \citet{Guth:2018} and \citet{Norton:2021} concluded that the notion of a uniform probability distribution on an infinite support is internally inconsistent, but we are not forced to draw this conclusion. From a logical viewpoint, we merely conclude that the assumptions are jointly inconsistent. Traditionally, probabilists have debated the issue as a problem of additivity, whereas the focus in the measure problem of cosmology was on non-normalizability.

Section~\ref{sec:AlternativeFormalisms} reviewed four ways of representing a fair lottery on the natural numbers, to wit: (1) FAP theory, (2) NAP theory, (3) Norton's ILL, and (4) non-normalizable quasi-probability theory. Physicists can help themselves to one or more of these approaches if the need arises.

As we have seen in section~\ref{sec:IFLparadoxinflation}, this need indeed presents itself in cosmology: the measure problem crucially depends on the interplay between infinite measures (and the limit processes involved) and probabilities. Moreover, there are at least two different ways in which uniform distributions take centre stage in cosmology. First, when a stationary measure is obtained via Liouville's theorem, it results in a uniform Lebesgue volume measure on the phase space. This measure is non-normalizable when the phase space is not compact. Second, a uniform measure over possible observers has been suggested in the cosmology literature, which again fails to be normalizable when there are infinitely many observers in the model of the universe or multiverse. In the probability literature, symmetry considerations are sometimes used to support uniform probability distributions (\textit{e.g.}, via Poincar\'e's method of arbitrary functions or de Finetti's representation theorem), but to the best of our knowledge these have not been applied yet in the cosmology literature.

Given that densities can be established for infinite lotteries, it seems at least possible that we might figure out how to model densities pertaining to inflationary pocket universes, covered in section~\ref{sec:measureprobleminflation}, too. Still, there is a crucial difference between the two cases. Whereas the natural numbers have a canonical order, that allows for a canonical choice for limit processes to obtain well-defined (alternative) probabilities, an important open question in inflationary cosmology is what the relevant structure is on the phase space to allow for an equally well-defined notion of probability in this context. In addition, it is not clear that a countably infinite lottery is a good model system to begin with, since fractal-like growth of pocket universes may result in an uncountable phase space.

Our review of various ways to model a fair lottery on the natural number does not directly help to answer a number of additional questions that arise in the context of the cosmological measure problem and that need to be analysed step by step, such as: how to factor in observer selection effects (which already occur in the finite case), how to apply typicality reasoning (in the sense of \citet{HartleSrednicki:2007} and \citet{GarrigaVilenkin:2008}; or whether this is useful in the first place), and how to confirm a theory of this type given that we use data from our only observable universe both for theory building and for confirmation \citep[double-counting, in the sense of][]{Friederich:2017}. Offering such a full analysis goes well beyond the scope of this contribution.

As a final remark, observe that our analysis of the infinite fair lottery paradox illustrates an approach for analysing paradoxes more generally---at least paradoxes that pose formal inconsistencies. The method relies on taking multiple combinations of assumptions (which may be axioms, properties, equations etc.) that are jointly consistent and putting the conclusions of these different fragments of the original problem together, to get a better sense of why combining all the assumptions was inconsistent, as well as the structure of a collection of neighbouring systems. The latter could be developed in \citeauthor{Gardenfors:2004}'s (\citeyear{Gardenfors:2004}) theory of conceptual spaces: it assumes a measure of similarity along various quality dimensions (defined by the nature of the assumptions at hand), which gives rise to a distance measure and thus the notion of a neighbourhood.

The situation is analogous to specifying equations for three mutually non-parallel lines in a plane expecting them to intersect in one point but finding that they intersect two-by-two in three points. Perhaps, ideally, there would be a common intersection point, but due to some practical errors in fact they do not---as happens, for instance, in imperfect constructions for linear perspective drawing. By investigating the location of these three `partial' intersections, one gets a sense of the region in which the three-way intersection could have been. Likewise, our analysis allowed us to investigate various sets of jointly consistent assumptions, each of which can be analysed without resulting in a paradox, which together suggest a way forward for the measure problems in cosmology.

It is worth repeating that this paper did not even cover all possible approaches: for instance, it did not explore qualitative probability theories \citep[such as][]{DiBella:2018}, Dempster--Shafer theory, or the option to represent the sample space by a hyperfinite set \citep{Nelson:1987}. Since ``probability is about modeling real world systems in order to understand and make predictions about that system'' \citep[p.~12]{Pfannkuch-etal:2016}, it seems wise to take a pluralist approach when dealing with various instances of the measure problem in cosmology: by studying the same questions in the context of different mathematical formalisms for probability, one can disentangle limitations of a particular framework from properties of the modelled system itself.

\section*{Funding}
This project was supported by the Research Foundation Flanders (FWO Grant No.\ G066918N).
\section*{Acknowledgements}
The results presented here were part of the essay that won the 2019 \textit{Philosophy of Cosmology Essay Contest} organized by the New Directions in Philosophy of Cosmology project; I am grateful to project directors Chris Smeenk and James Owen Weatherall and all involved for the encouragement. I also thank three anonymous reviewers of this journal, whose reports helped me to improve this paper.


\bibliographystyle{cas-model2-names}

\bibliography{cas-refs}

\end{document}